\documentclass[manuscript]{aastex631}

\usepackage{chngcntr}
\usepackage{appendix}
\usepackage{natbib}
\usepackage{CJK}
\usepackage{longtable}
\usepackage{amsmath}
\usepackage{multirow}
\usepackage{subfigure}
\usepackage{graphicx}
\usepackage{txfonts}
\usepackage{epsfig}
\usepackage{color,xcolor}
\usepackage{multirow}
\usepackage{array}
\usepackage{soul}
\newcolumntype{P}[1]{>{\centering\arraybackslash}p{#1}}
\usepackage{longtable}

\shorttitle{CS in the 21\,$\mu$m sources}
\shortauthors{Qiu et al.}
\begin{document}
\begin{CJK*}{UTF8}{gbsn}

\title{
Gas-phase Molecules in Protoplanetary Nebulae with the 21\,$\mu$m Emission Feature\\  
\uppercase\expandafter{\romannumeral2}. 
Carbon monosulfide
}

\correspondingauthor{Yong Zhang}
\email{zhangyong5@mail.sysu.edu.cn}

\author[0000-0002-9829-8655]{Jian-Jie Qiu (邱建杰)}
\affiliation{School of Mathematics and Physics, Jinggangshan University, 28 Xueyuan Road, Qingyuan District, Ji'an 343009, Jiangxi Province, China}
\affiliation{Key Laboratory of Energy Conversion Optoelectronic Functional Materials of Jiangxi Education Institutes, Ji'an, 343009, China}
\affiliation{Institute for Astronomy and Astrophysics, Department of Physics, JingGangShan University, Ji'an, 343009, China}

\author[0000-0002-1086-7922]{Yong Zhang (张泳)}
\affiliation{School of Physics and Astronomy, Sun Yat-sen University, 2 Da Xue Road, Tangjia, Zhuhai 519000, Guangdong Province, China}
\affiliation{Xinjiang Astronomical Observatory, Chinese Academy of Sciences, Urumqi, 830011, Xinjiang Province, China}
\affiliation{CSST Science Center for the Guangdong-Hongkong-Macau Greater Bay Area, Sun Yat-Sen University, Zhuhai, China}
\affiliation{Laboratory for Space Research, The University of Hong Kong, Hong Kong, China}

\author{Deng-Rong Lu (逯登荣)}
\affiliation{Purple Mountain Observatory, Chinese Academy of Sciences, Nanjing 210034, Jiangsu Province, China}

\author[0000-0002-5981-7846]{Zheng-Xue Chang (常正雪)}
\affiliation{College of Mathematics and Physics, Handan University, Handan, 056005, Hebei Province, China}

\author[0000-0002-5161-8180]{Jiang-Shui Zhang (张江水)}
\affiliation{Center For Astrophysics, Guangzhou University, 230 Wai Huan Xi Road, Guangzhou Higher Education Mega Center, Guangzhou 510006, Guangdong Province, China}

\author[0000-0003-2090-5416]{Xiao-Hu Li (李小虎)}
\affiliation{Xinjiang Astronomical Observatory, Chinese Academy of Sciences, Urumqi, 830011, Xinjiang Province, China}

\author[0000-0002-4154-4309]{Xin-Di Tang (汤新弟)}
\affiliation{Xinjiang Astronomical Observatory, Chinese Academy of Sciences, Urumqi, 830011, Xinjiang Province, China}

\author[0000-0002-7716-1094]{Yisheng Qiu (邱逸盛)}
\affiliation{Research Center for Astronomical Computing, Zhejiang Laboratory, Hangzhou, China}

\author[0000-0003-3324-9462]{Jun-ichi Nakashima (中岛淳一)}
\affiliation{School of Physics and Astronomy, Sun Yat-sen University, 2 Da Xue Road, Tangjia, Zhuhai 519000, Guangdong Province, China}
\affiliation{CSST Science Center for the Guangdong-Hongkong-Macau Greater Bay Area, Sun Yat-Sen University, Zhuhai, China}

\author{Lan-Wei Jia (贾兰伟)}
\affiliation{Center For Astrophysics, Guangzhou University, 230 Wai Huan Xi Road, Guangzhou Higher Education Mega Center, Guangzhou 510006, Guangdong Province, China}

\begin{abstract}

The carrier of the 21\,$\mu$m emission feature discovered in a handful of protoplanetary nebulae (PPNe) is one of the most intriguing enigmas in circumstellar chemistry. 
Investigating the gas-phase molecules in PPNe could yield important hints for understanding the 21\,$\mu$m feature. 
In this paper, we report observations of the CS $J=5 \to 4$ line at 245\,GHz and the CO $J=1 \to 0$ line at 115\,GHz toward seven PPNe exhibiting the 21\,$\mu$m feature. 
We find that CS is extremely scarce in these PPNe and the CS line is only detected in one source, IRAS\,Z02229$+$6208. 
Based on the assumption of local thermal equilibrium and negligible optical depth, we derive that the CS column densities and fractional abundances relative to H$_{2}$ are $N({\rm CS}) < 9.1\times10^{13}$\,cm$^{-2}$ and $f({\rm CS}) < 8.1\times10^{-7}$. 
A comparison of the CS abundances across different circumstellar envelopes reveals that the variations in CS abundance are complex, depending not only on the evolutionary stages but also on the properties of individual objects.

\end{abstract}


\section{Introduction} \label{sec:intro}

When low- and intermediate-mass (0.8--8.0\,$M_{\odot}$) stars transition from the asymptotic giant branch (AGB) stage and evolve toward the planetary nebula (PN) stage, they undergo a protoplanetary nebula (PPN) stage \citep{Herwig05}, during which massive material loss has terminated and the central star is not hot enough to emit sufficient ultraviolet (UV) photons to ionize the circumstellar envelopes (CSEs) \citep{Kwok93}. 
During this short-lived PPN stage ($\sim$10$^{3}$\,yr), the envelope shape changes from roughly spherically symmetric to a large diversity of morphologies \citep{Balick02}, and the chemical compositions significantly alter \citep{Cernicharo11}. 
So far, more than 100 gas-phase molecules have been detected in the CSEs of AGB stars \citep{Cernicharo11, Endres16, Decin21, Tuo24}. 
Complex organic compounds (COMs) such as aromatic and aliphatic organic nanoparticles, which are regarded as the carriers of the Unidentified Infrared Emission bands, have been predominantly detected in PPNe. 
This observation might indicate that COMs can be efficiently synthesized during the PPN phase \citep{Kwok11, Kwok24b}, although alternative explanations cannot be ruled out\footnote{
For instance, COMs could form in the AGB phase but remain undetectable until they are irradiated in the PPN phase \citep[e.g.,][]{Speck97}. 
Additionally, aromatic molecules can be released when amorphous carbon dust is disrupted during the PPN phase.
}. 
With the CSE's expansion, parts of the circumstellar materials may survive and eventually disperse into the interstellar medium (ISM), seeding the chemistry in the formation regions of the next-generation stellar and planetary systems \citep{Kwok22}. 
Therefore, investigating the molecules in PPNe could provide significant clues for our understanding of the lifecycle of galactic materials.

The 21\,$\mu$m feature is an unidentified emission feature appearing in the infrared (IR) spectra of a small fraction of PPNe \citep[see][for a recent review]{Volk20}. 
It was first discovered by \citet{Kwok89} in the Low Resolution Spectrometer spectra of four \textit{Infrared Astronomical Satellite (IRAS)} sources. 
Based on higher spectral resolution observations from the Short-Wavelength Spectrometer aboard the \textit{Infrared Space Observatory}, \citet{Volk99} demonstrated that this feature exhibits an asymmetric profile. 
Its peak wavelength, where the intensity reaches a maximum, is 20.1\,$\mu$m, with a width of 2\,$\mu$m. 
In contrast, its central wavelength, representing the midpoint of the profile, was determined to be $20.47\pm0.10$\,$\mu$m \citep{Sloan14}.

To date, the 21\,$\mu$m feature has been detected only in 31 objects, including 20 in the Milky Way and eleven in the Magellanic Clouds; all are carbon-rich \citep{Volk20}. 
Given the rarity of the 21\,$\mu$m feature, \citet{Kwok89} proposed that it is likely a transient phenomenon during the PPN stage. 
Subsequent studies have generally assumed this emission feature to be associated with dust grains \citep{Hrivnak99, Volk99}. 
To date, however, it remains unclear whether this feature corresponds to a solid-state dust feature or a free-flying molecular emission band \citep{Volk20}. 
\cite{Mishra16} performed a statistical study of 18 Galactic 21\,$\mu$m sources and found that the mass-loss rate in their AGB phase is strongly correlated to the 21\,$\mu$m flux. 
They thus suggested that the 21\,$\mu$m carrier may be formed during the AGB phase but not emit until the PPN phase. 
Identification of the 21\,$\mu$m carrier is essential for refining the state-of-art chemical network and understanding circumstellar chemistry. 
However, accurate identification of the 21\,$\mu$m carrier is still a long-unresolved debate. 
Up to now, the proposed candidates, including SiS$_{2}$, Fe$_{\rm m}$O$_{\rm n}$, TiC, thiourea, hydrogenated fullerenes, polycyclic aromatic hydrocarbons, hydrogenated amorphous carbon, nano-diamonds, etc. (see \citealt{Volk20} for an overview), have exceeded the number of detected 21\,$\mu$m sources.

Since both the gas-phase chemistry and dust process interactively depend on physical environments and chemical compositions in the CSEs, the study of gas-phase molecules could provide some clues and constraints for the identification of the 21\,$\mu$m carrier. 
Through a molecular line survey toward a 21\,$\mu$m source IRAS\,22272$+$5435 and a non-21\,$\mu$m source IRAS\,21318$+$5631 at the 3 and 1.3\,mm bands, \cite{Zhang20} found that the 21\,$\mu$m source shows relatively stronger emission lines from SiC$_{2}$ and HC$_{3}$N but weaker ones from SiS. 
Through a study of six 21\,$\mu$m sources, \citet{Qiu24} (Paper~I) found that the relative intensities of the 21\,$\mu$m feature with respect to total IR emission have weak or no correlations with the fractional abundances of gas-phase molecules. 
More observations are badly required to draw more meaningful conclusions.

All the 21\,$\mu$m sources come with a broad emission feature at 30\,$\mu$m that is generally attributed to MgS \citep{Volk20}. 
Therefore, it is reasonable to speculate that the 21\,$\mu$m carrier may have some connection with sulfur chemistry. 
The abundances of sulfur in PNe have been found to be systematically lower than those in \ion{H}{2} regions and blue compact galaxies with the same metallicity. 
This is the so-called `sulfur anomaly' issue in PN study, first identified by \cite{Henry04} and confirmed by \cite{Henry12}. 
\cite{Henry04} ascribed this to the gross underestimating of the proportion of element S in the S$^{+3}$ format. 
However, using argon as a superior PN metallicity indicator, \cite{Tan24} found that the deficiency of sulfur is mainly in carbon-rich PNe with an intermediate progenitor mass ($\sim$4\,$M_{\odot}$) and dispelled the hypotheses that the `sulfur anomaly' originates from an underestimation of higher sulfur ionization stages. 
They attributed the sulfur deficit to efficient dust formation in these objects. 
Gas-phase sulfur could be substantially incorporated into pre-existing dust grains and form a coating. 
As the direct progenitor of PNe, PPNe serve as an excellent site to investigate the adsorption process of gas-phase sulfur. 
However, it should be borne in mind that there is a bias when conducting a comparative analysis between the abundance of gas-phase sulfur in PPNe and PNe. 
The sulfur abundance in PNe can be easily determined via forbidden lines, whereas detecting sulfur-bearing molecules in PPNe is more challenging. 
Therefore, those PPNe in which sulfur molecules have been detected may not represent typical carbon-rich objects.

Carbon Monosulfide (CS) is an abundant molecule in carbon-rich AGB envelopes. 
In the carbon-rich PPN CRL\,2688, which is speculated to have just exited the 21\,$\mu$m feature stage \citep{Geballe92}, CS emission lines are stronger than those of other sulfur-bearing molecules \citep{Park08, Zhang13, Qiu22}. 
While \cite{Geballe92}'s speculation lacks observational support and CRL\,2688 may not represent a typical carbon-rich PPN, we nonetheless select CS as a tracer for gas-phase sulfur in 21\,$\mu$m sources. 
Non-equilibrium chemical models suggest that the formation of CS primarily occurs via the reaction S + CN $\longrightarrow$ CS + N, where CN is derived from the photodissociation of HCN \citep{Cherchneff06}.

In this work, we report new observations of the CS $J=5 \to 4$ transition at the rest frequency of 244935.556\,MHz toward seven 21\,$\mu$m sources. 
It is the second part of the series that presents our study of gas-phase molecules in 21\,$\mu$m sources. 
This paper is organized as follows. 
In Sect.~2, we introduce the observations and data reductions. 
The observational results, column densities, and fractional abundances are presented in Sect.~3. 
In Sect.~4, we discuss the implications of our observations on circumstellar chemistry. 
A summary is given in Sect.~5.

\section{Observations and Data reduction}

The sample contains seven PPNe exhibiting the 21\,$\mu$m feature, 
IRAS\,Z02229$+$6208, IRAS\,05341$+$0852, IRAS\,06530$-$0213, IRAS\,07134$+$1005, IRAS\,22223$+$4327, IRAS\,22272$+$5435, and IRAS\,23304$+$6147, 
which were selected based on their strong CO emission.
The CS $J=5 \to 4$ transition observations at the 1.3\,mm band were carried out with the Arizona Radio Observatory 10\,m Submillimeter Telescope (SMT)\footnote{https://aro.as.arizona.edu/?q=facilities/uarizona-aro-submillimeter-telescope} at Mt.~Graham on 2022 November 28 and 2023 October 29--31. 
The 1.3\,mm receiver utilizes ALMA Band 6 sideband-separating (SBS) dual-channel Superconductor Insulator Superconductor (SIS) mixers, with image rejection typically about 18\,dB. 
The 1024-channel acousto-optical spectrometer and the 512-channel Forbes Filterbanks were used simultaneously with a channel width of 1\,MHz. 
The corresponding velocity resolution of the CS $J=5 \to 4$ line is about 1.22\,km\,s$^{-1}$. 
The position-switching mode was employed with an azimuth offset of $\pm$2$\arcmin$. 
Pointing and focus were checked every 2 hours using nearby planets or strong continuum sources. 
The on-source time of each scan is 6 minutes. 
The half-power beam width (HPBW) is about 30$\rlap{.}\arcsec$8 at 250\,GHz. 
The pointing offset is less than 10$\arcsec$. 
Thus, the effect of pointing offset on the measurements could be neglected. 
The main beam temperature ($T_{\rm mb}$) is determined by the expression of $T_{\rm mb} = T_{A}^{*} / \eta_{\rm mb}$, where $T_{A}^{*}$ and $\eta_{\rm mb}$ are the antenna temperature and main beam efficiency of 0.7 at the SMT 1.3\,mm band, respectively. 
The typical system temperatures are about 200--300\,K in $T_{\rm mb}$ scale.

The observations of the CO $J=1 \to 0$ transition in the 3\,mm window were conducted utilizing the Purple Mountain Observatory (PMO) 13.7\,m millimeter telescope\footnote{http://dlh.pmo.cas.cn/wyjsbjs/hmbwyj/} at Delingha in China on 2024 June 1 and 2025 March 14--29. 
Two sidebands SIS receiver was used. 
Spectrometer backends for each sideband utilize a Fast Fourier Transform Spectrometer (FFTS) with a bandwidth of about 1\,GHz including 16384 channels and a frequency resolution of 61\,kHz. 
To increase the signal-to-noise ratio, we combined the data of \cite{Qiu24} and the 3\,mm spectra were smoothed over several channels using the `boxcar' method to obtain a rebinned frequency resolution close to that of 1.3\,mm spectra. 
The final velocity resolution of the CO $J=1 \to 0$ transition is about 1.1\,km\,s$^{-1}$. 
The main beam efficiency of the PMO 13.7\,m telescope is 0.5 at the 3\,mm band. 
The HPBW is about 52$\arcsec$ at 115 GHz. 
Pointing and focus were checked every two hours.

All spectral data were reduced by using the Continuum and Line Analysis Single-dish Software (CLASS) package of Grenoble Image and Line Data Analysis Software (GILDAS)\footnote{https://www.iram.fr/IRAMFR/GILDAS/} \citep{Pety05}. 
We inspected each spectrum and discarded those with a significantly larger noise level than theoretical expectations due to bad atmospheric conditions and receiver instabilities. 
A first-order baseline subtraction was performed for each spectrum. 
We co-added all spectra at the same frequency coverages to improve the signal-to-noise ratio, using the reciprocal value of the noise level as the weight.

We fitted the line profiles to derive the velocity-integrated intensities using a stellar-shell model with the expression of 
\begin{equation} 
f(\nu) = \frac{A}{\Delta\nu} \frac{1+4H[(\nu-\nu_{0})/\Delta\nu]}{1+H/3}. 
\label{Equation1}
\end{equation}  
A detailed description of the stellar-shell model and the parameters in Eq.~(1) can be found in Paper~I. 
The derived results, including expansion velocity ($V_{\rm exp}$), Horn/Center parameter ($H$), peak intensity in main beam temperature scale ($T_{\rm p}$), velocity-integrated intensity ($\int T_{\rm mb}{\rm d}V$), root mean square (RMS) noise level of the baseline, and line-center velocity ($V_{\rm LSR}$) are listed in Table~\ref{Table1}.

\section{Results} \label{sec:style}

\begin{figure*}
\epsscale{.50}
\plotone{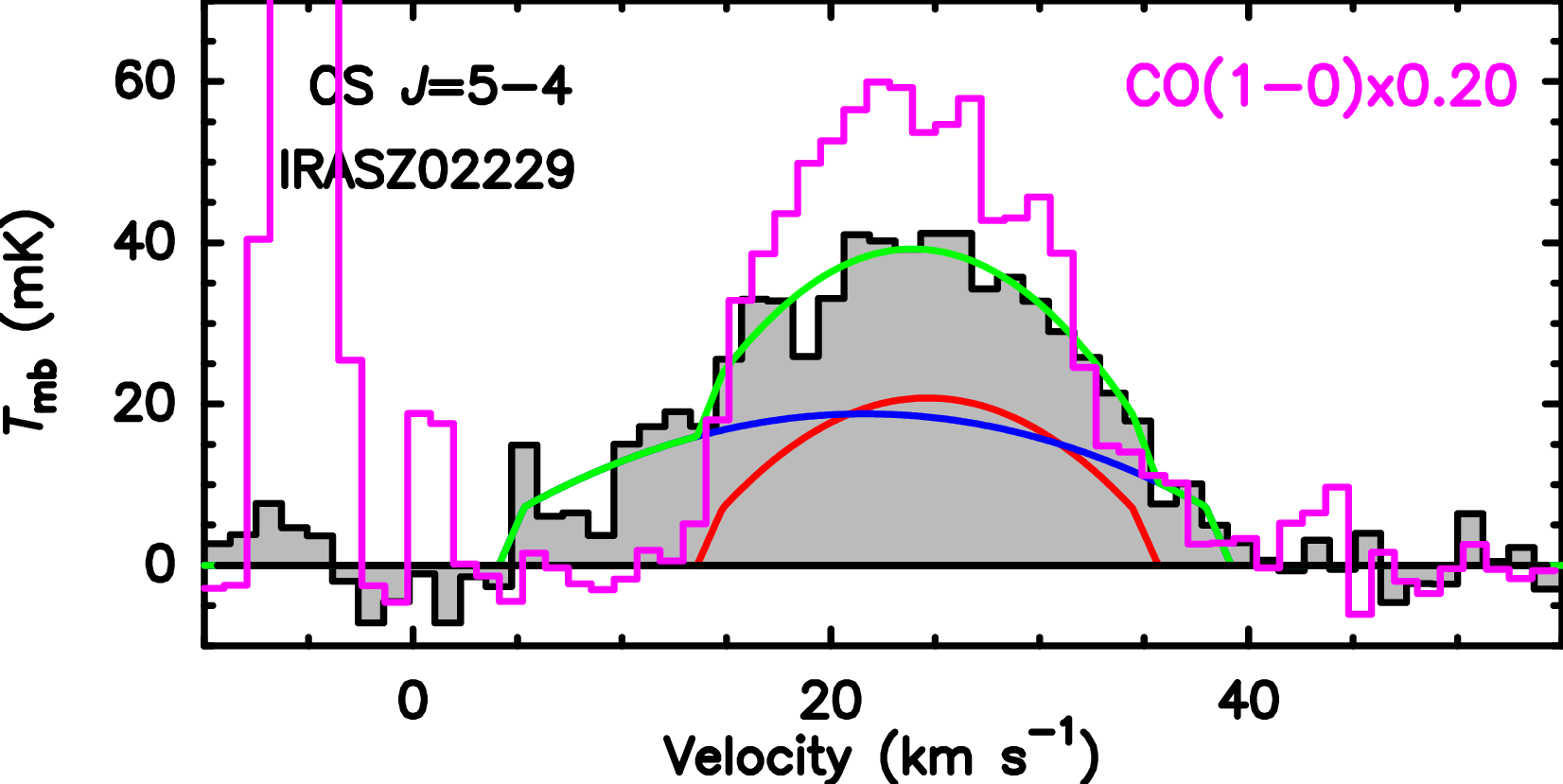}
\plotone{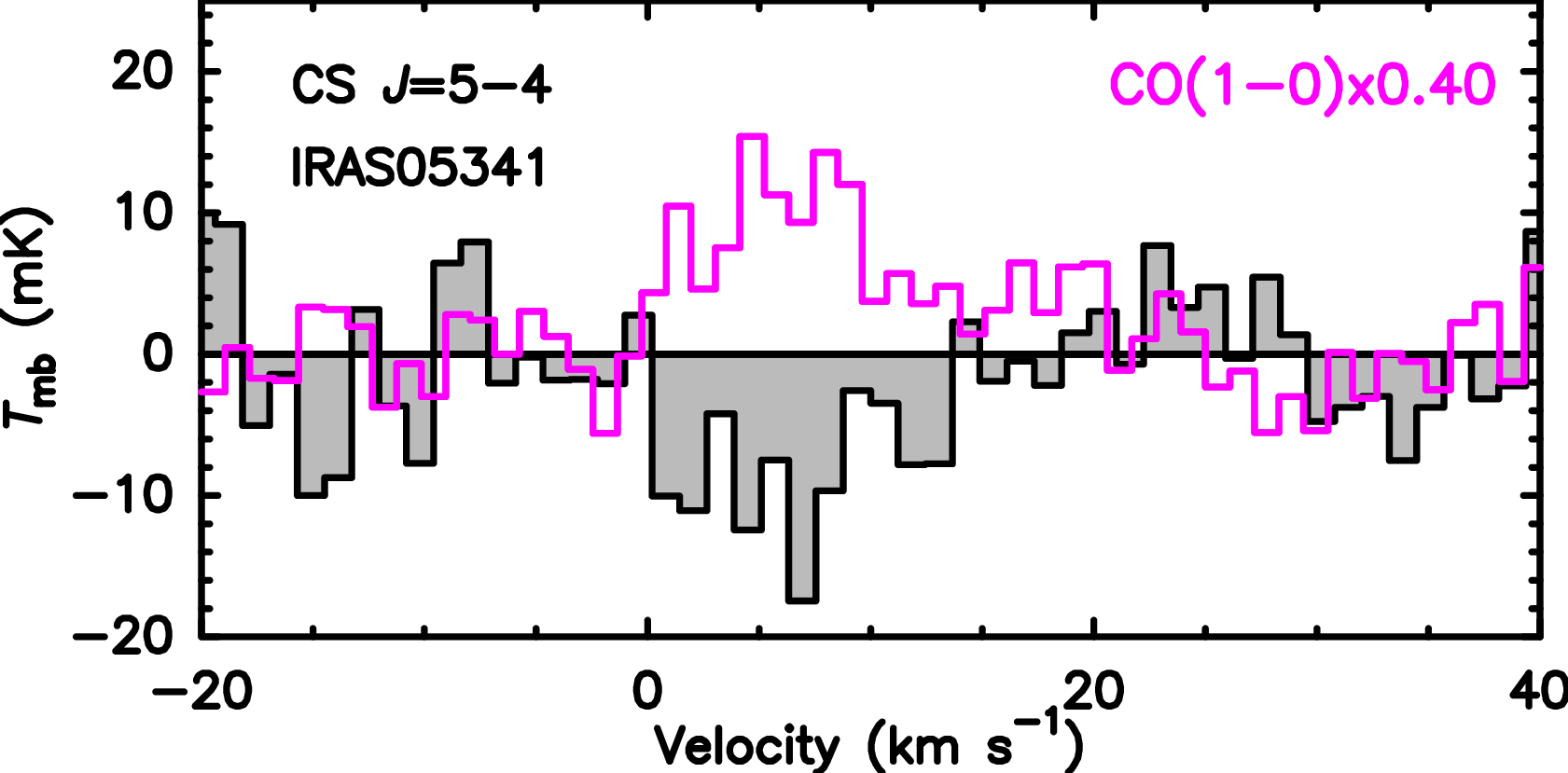}
\plotone{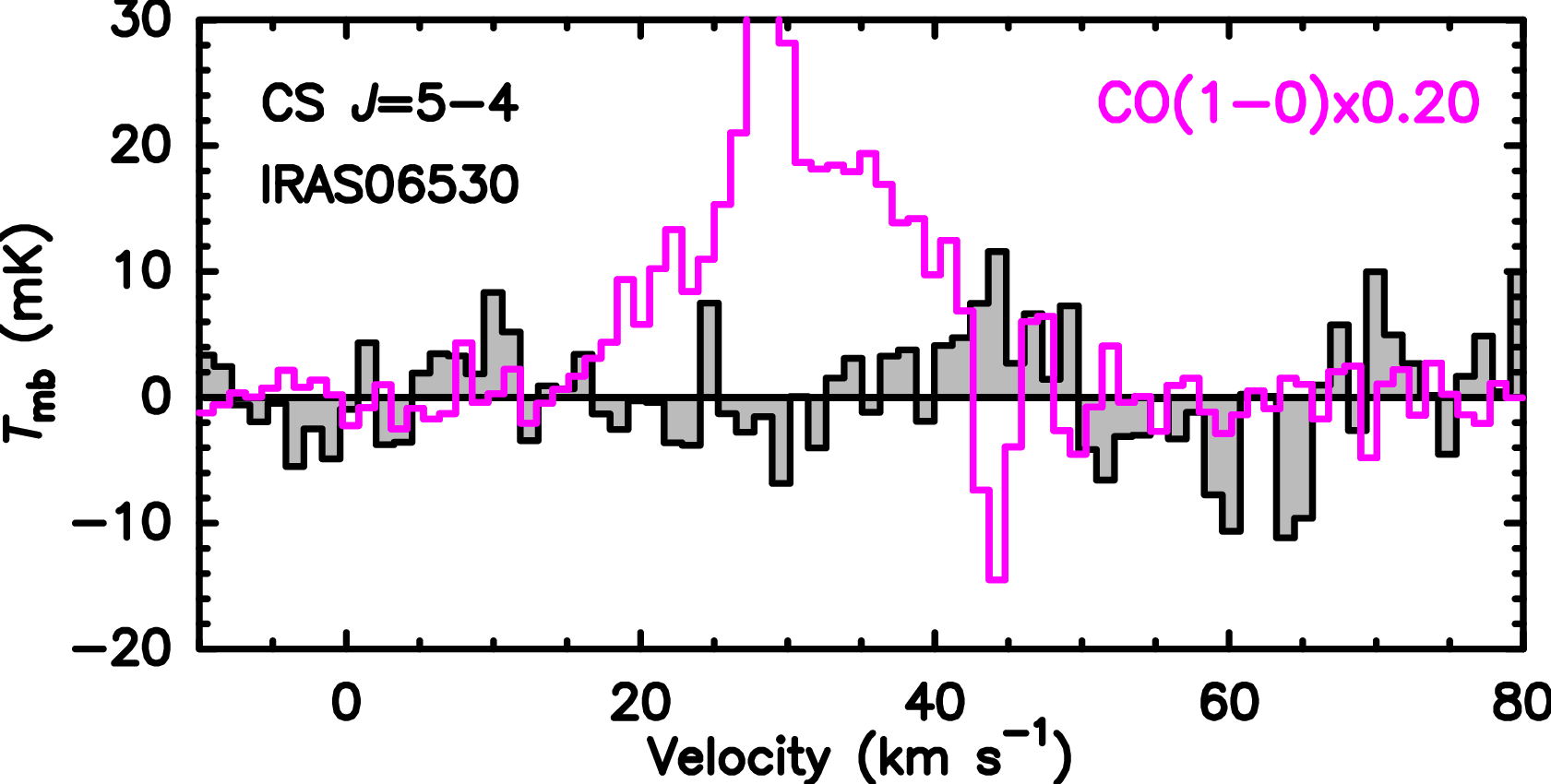}
\plotone{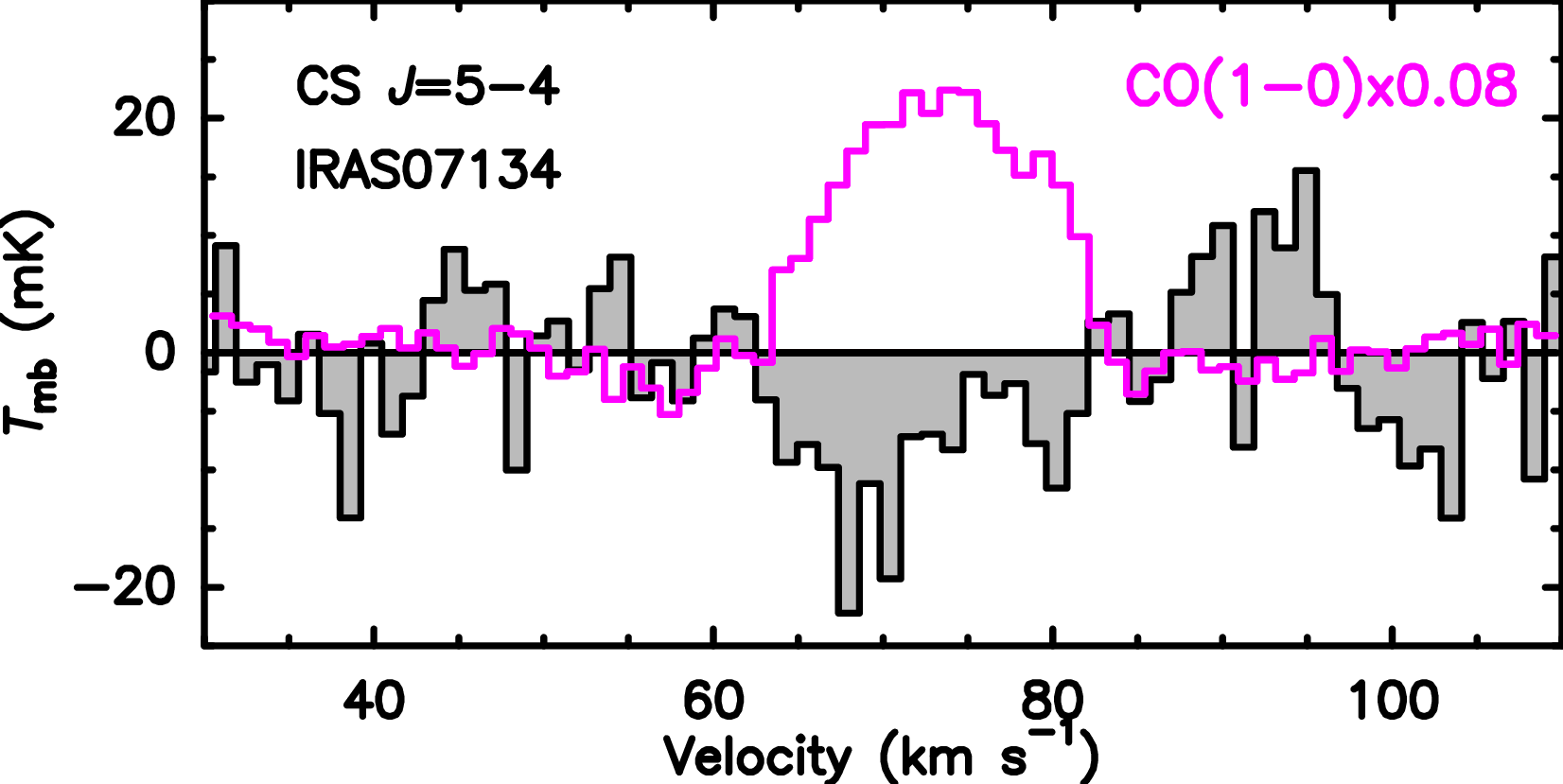}
\plotone{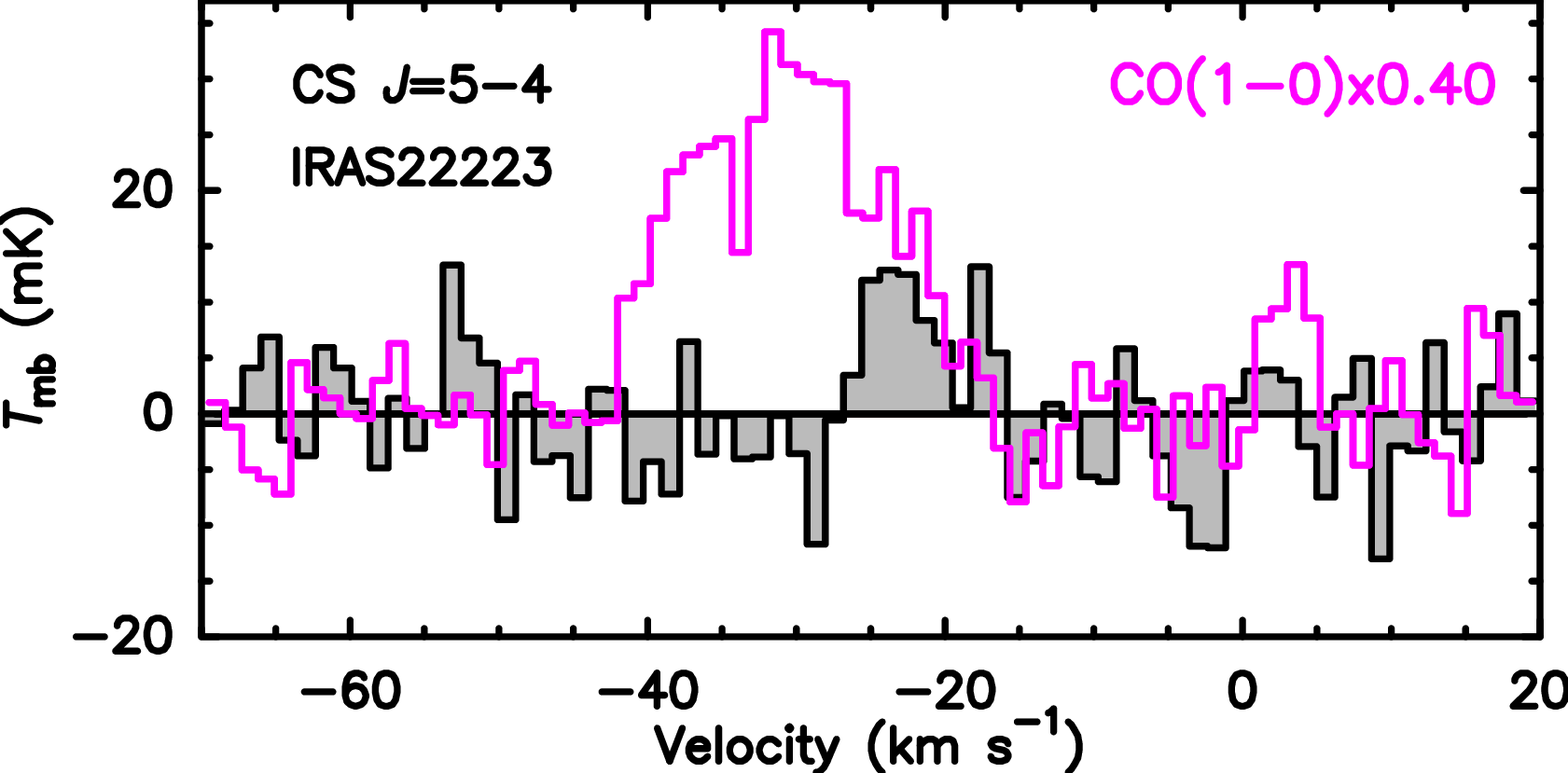}
\plotone{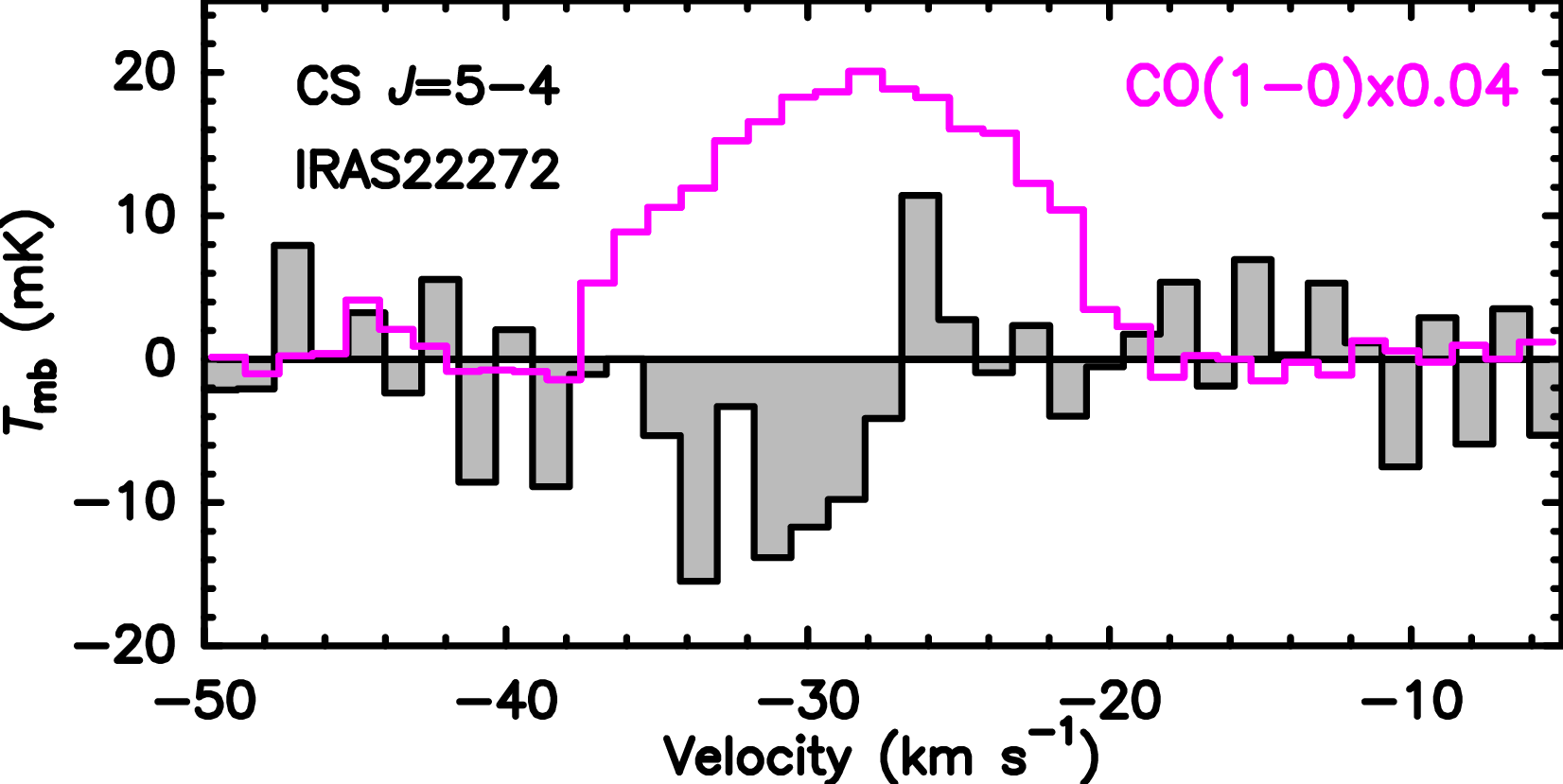}
\plotone{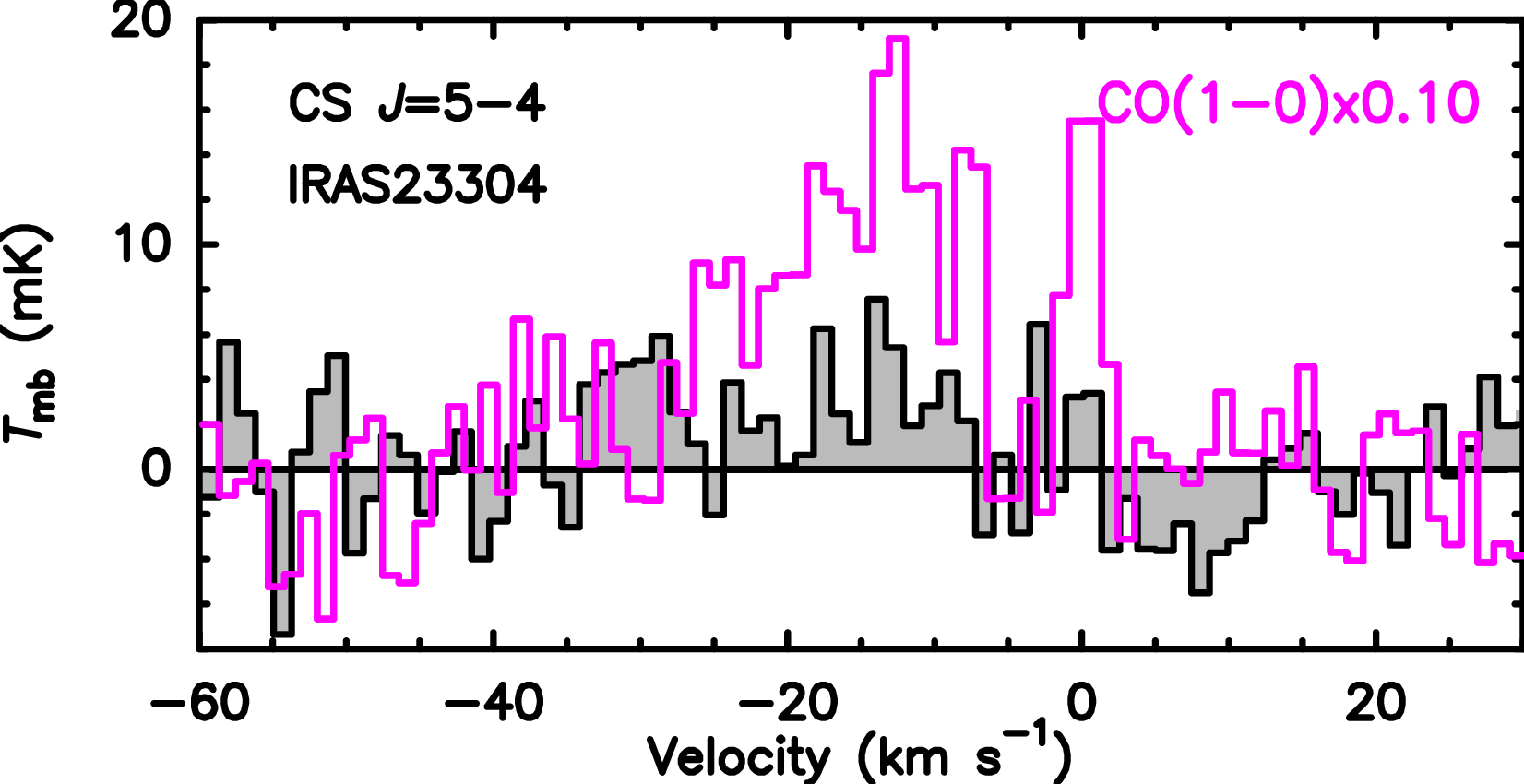}
\caption{Spectra of CS $J=5 \to 4$ (filled black curves) and CO $J=1 \to 0$ (magenta curves). 
The red, blue, and green curves represent the stellar-shell fitting of the narrow component, broad component, and total profile, respectively. 
}
\label{Figure1}
\end{figure*}

The CS $J=5 \to 4$ transitions overlaid with the CO $J=1 \to 0$ transitions are shown in Fig.~\ref{Figure1}. 
IRAS\,Z02229$+$6208 is the only source in which the CS line is evidently detected. 
For the other sources, we estimate the upper limits of the CS line strengths based on the noise levels (1$\sigma$ $\sim$ 5\,mK). 
In previous literature, the detections of CO, $^{13}$CO, HCN, H$^{13}$CN, HNC, CN, and HC$_{3}$N have been reported in IRAS\,Z02229$+$6208 \citep{Hrivnak99, Hrivnak05, Qiu24}. 
Our detection of CS adds a new molecule to the chemical inventory of this source. 
The CS line profile shows a broad line wing. 
We thus fit this line using a two-component stellar-shell model. 
In IRAS\,Z02229$+$6208, the $J=1 \to 0$ and $2 \to 1$ transitions of CO and $^{13}$CO and the $J=3 \to 2$ transition of HCN and H$^{13}$CN also exhibit a broad wing \citep{Hrivnak05, Qiu24}.

\subsection{Column Densities and Fractional Abundances}

Under the local thermal equilibrium (LTE) assumption, we derive the column densities ($N$) of CO and CS through the level population expression 
\begin{equation}
N = \frac{8\pi k \, \nu^{2} \, Q(T_{\rm ex})}{hc^{3} \, A_{\rm ul} \, g_{\rm u}} \, e^{E_{\rm u}/kT_{\rm ex}} \, C_{\tau} \int T_{\rm S} \, {\rm d}v ,
\label{Equation2}
\end{equation} 
where $k$ and $h$ are the Boltzmann constant and Planck constant, respectively, 
$\nu$ is the rest frequency of the transition in Hz, 
$E_{\rm u}/k$ is the upper-level energy in K, 
and $Q(T_{\rm ex})$, $A_{\rm ul}$, and $g_{\rm u}$ are the partition function, spontaneous emission coefficient, and upper state degeneracy, respectively, taken from the CDMS catalog \citep{Muller01, Muller05, Endres16}. 
$\int T_{\rm S}\,{\rm d}v$ is the velocity-integrated intensity, 
where $T_{\rm S}$ is the source brightness temperature of the emission line. 
Under the assumption of a Gaussian distribution of the surface brightness, 
the conversion between $T_{\rm S}$ and $T_{\rm mb}$ follows the expression of 
$T_{\rm S}=T_{\rm mb}(\theta_{\rm b}^{2} + \theta_{\rm s}^{2})/\theta_{\rm s}^{2}$, 
where $\theta_{\rm b}$ is the antenna HPBW of the emission line.  
The angular size of the source, $\theta_{\rm s}$, is derived from the outer radius of the shell and the distance given in \cite{Mishra16}\footnote{
The modeled radii of the nebulae were constrained by observed optical and infrared morphologies presented in angular scale. Thus, in estimating $\theta_{\rm s}$, we did not endeavor to apply corrections based on the updated distances.
}.
$C_{\tau}$ = $\tau / (1-e^{-\tau})$ is an optical-depth ($\tau$) correction factor. 
Given that most of the CS $J=5 \to 4$ lines are not detected, we thus ignore the influence of optical depth ($C_{\tau} = 1$). 
The CO $J=1 \to 0$ lines show parabolic profiles, indicating optically thick emission. 
The calculated column densities of CO are the lower limits if we do not consider the influence of optical depth. 
Assuming $\tau$ = 1, $N$ increases by 50\% compared to the optically thin case. 
In previous literature \citep{Zhang20,Qiu22}, the excitation temperatures of CS in IRAS\,21318$+$5631, IRAS\,22272$+$5435, and CRL\,2688 are 15, 17, and 13\,K, respectively. 
Therefore, we assume an excitation temperature of 15\,K for our calculations.
It should be noted that $N$ is insensitive to $T_{\rm ex}$. 
For example, $N(\rm CS)$ would decrease by about 24\% if assuming $T_{\rm ex}$ = 80\,K. 
The calculated results are listed in Table~\ref{Table1}.

To determine the fractional abundances ($f_{\rm X}$) with respect to H$_{2}$, 
we utilize the equation proposed by \cite{Olofsson97}, 
\begin{equation}
f_{\rm X} = 1.7 \times 10^{-28} \frac{V_{\rm exp} \theta_{\rm b}D}{\dot{M}_{\rm H_2}} \frac{Q(T_{\rm ex})\nu^{2}}{g_{\rm u}A_{\rm ul}} \frac{e^{E_{\rm l}/kT_{\rm ex}}\int T_{\rm mb}{\rm d}V}{\int^{x_e}_{x_i}e^{-4x^2{\rm ln2}}{\rm d}x},
\label{Equation3}
\end{equation}  
where $E_{\rm l}$ is the energy of lower level, $D$ is the distance, $\dot{M}_{\rm H_{2}}$ is the H${_2}$ mass-loss rate, and $x_{i}$ and $x_{e}$ represent the ratios of the inner ($R_{i}$) and outer ($R_{e}$) radii over the beam size, respectively (i.e., $x_{e,i} = R_{e,i}/(\theta_{\rm b}D)$). 
The distances of the sources are derived from \textit{Gaia} DR3 \citep{Gaia23}. 
The expansion velocity of $V_{\rm exp}$ is taken from \cite{Mishra16}. 
The values of $\dot{M}_{\rm H_{2}}$, $R_{i}$, and $R_{e}$ provided in \cite{Mishra16} have been corrected using the \textit{Gaia} DR3 distances. 
The obtained $f_{\rm X}$ values are listed in the last column of Table~\ref{Table1}. 
The overall uncertainties in our derived results are predominantly determined by those associated with $\dot{M}$, which are estimated to be approximately 50\% \citep{Mishra16}.

\subsection{\texorpdfstring{Temporal Variation of the CS $J=5 \to 4$ Transition in IRAS\,22272$+$5435}{}}

The CS $J=5 \to 4$ transition in IRAS\,22272$+$5435 has been detected by \cite{Zhang20} utilizing the SMT-10\,m telescope in January 2009, which has $T_{\rm peak}$ of $\sim0.1$\,K. 
However, our observations using the same telescope in October 2023 show $T_{\rm peak}<0.01$\,K for this line. 
The discrepancy in flux cannot be accounted for by measurement uncertainties. 
\cite{Ueta01} hypothesized that IRAS\,22272$+$5435 might have experienced a sudden mass ejection in 1990. 
The latest \textit{UVB} multicolor photometry studies completed by \cite{Ikonnikova25} with data spanning over 30 years (1991--2024) suggested that IRAS\,22272$+$5435 may be a binary system, where shock waves induced by pulsational activity could trigger dust condensation. 
One might hypothesize that CS could be depleted into newly formed dust grains. 
However, we cannot find any subsequent mid-IR imaging or spectroscopic observations supporting the formation of new dust or an increase in the mass-loss rate. 
\cite{Hrivnak94} observed that the first overtone CO band transitioned from emission to absorption over a three-month interval, potentially related to stellar pulsations. 
Nevertheless, the line width of the CS line indicates an origin in the CSE far from the stellar atmosphere. 
Thus, the CS line variability likely arises from a distinct process compared to the CO IR band variability, with a presumably longer timescale. 
Consequently, we refrain from drawing definitive conclusions regarding the cause of the CS line's peculiar behavior. 
Additionally, we note that variability in the CS $J=1 \to 0$ transition has been detected toward the carbon star IRC$+$10216 \citep{Chau12}. 
Such a behavior bears a resemblance to the characteristics of a maser.

\section{Discussion}
\label{Section4}

\begin{figure*}
\epsscale{1.10}
\plotone{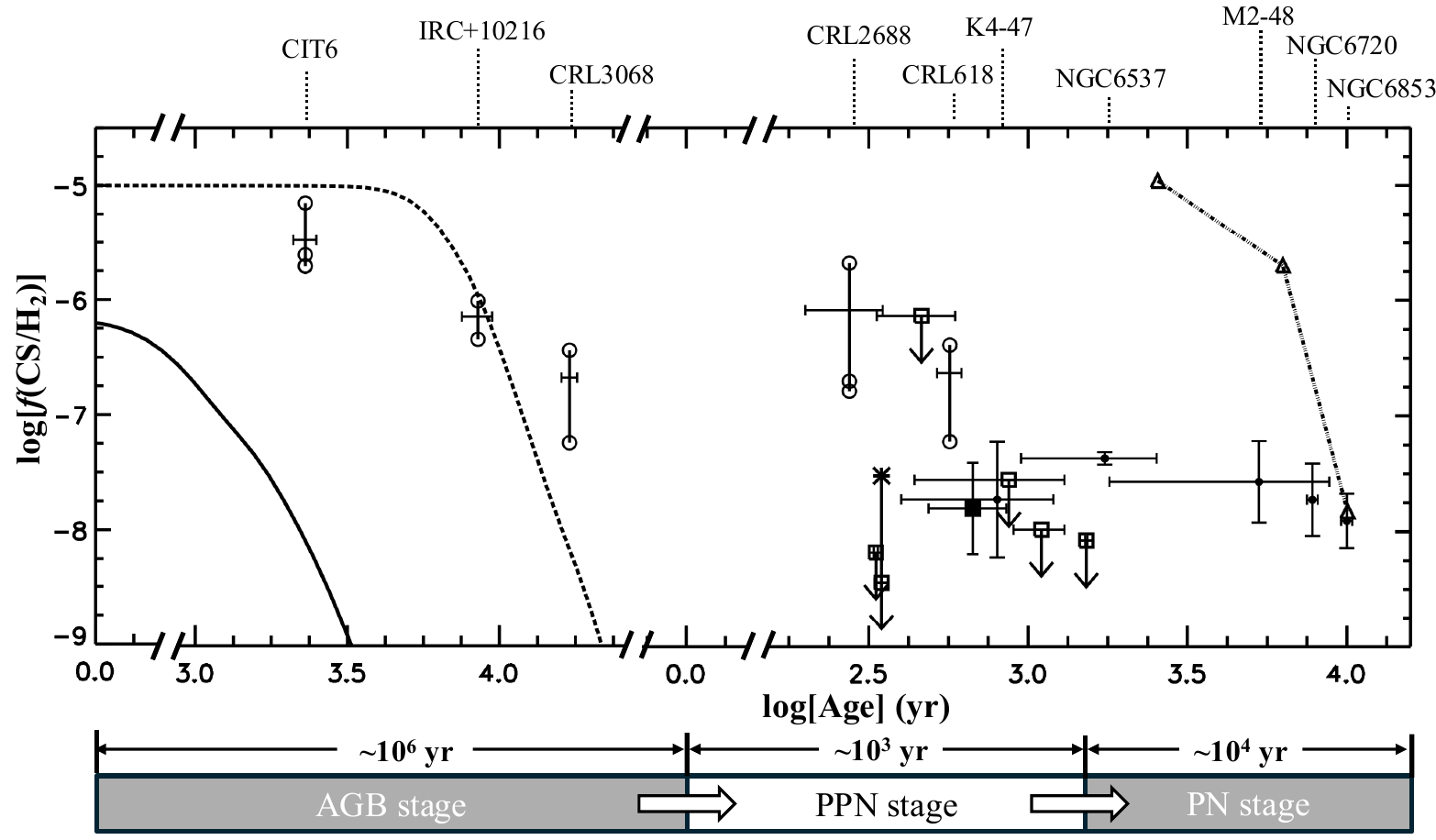}
\caption{
Fractional abundance of CS as a function of the dynamical age of the CSEs. 
The filled and open squares denote the 21\,$\mu$m sources with and without CS detection, respectively. 
These data points have been artificially shifted slightly for the sake of clarity. 
The filled circles denote the PNe investigated by \cite{Edwards14}. 
The open circles linked by straight lines denote the AGB CSEs and PPNe investigated in the literature listed in Table~\ref{Table2}. 
The solid and dashed curves represent the calculated results from Model~1 and Model~2 (see the text), respectively. 
The open triangles linked by dashed-dotted line are the calculated results of the clumpy model for PNe at the dynamical ages of 2\,550, 6\,300, and 10\,050\,yr \citep{Redman03}. 
The vertical dashed lines mark the dynamical ages and source names of the comparison sample. 
The asterisk denotes the fractional abundance of CS in IRAS\,22272$+$5435 calculated from the observations reported in \cite{Zhang20}, and this value is connected to our results by a straight line for visual comparison. 
}
\label{Figure2}
\end{figure*}

The evolution of CS in CSEs is not fully understood. 
\cite{Gold24} found that the CS abundance in PNe does not significantly vary with age across about 10\,000\,yr, while chemical models predict a dramatic decreasing trend. 
CS is very abundant in AGB envelopes. 
To investigate the CS evolution in the brief PPN stage, we thus plot in Fig.~\ref{Figure2} the fractional abundances of CS as a function of dynamical age. 
We select CIT\,6, IRC$+$10216, and CRL\,3068 in the AGB stage, CRL\,2688 and CRL\,618 in the PPN stage, and K4-47, NGC\,6537, M2-48, NGC\,6720, and NGC\,6853 in the PN stage as the comparison sample. 
The CS abundances of the comparison sample are taken from the references, as listed in Table~\ref{Table2}. 
The dynamical ages of the CSEs are derived based on their sizes and expansion velocities (see Appendix~\ref{AppendixA} for the details). 
The 21\,$\mu$m sources have similar dynamical ages, although the precise values are unknown. 
We simply assume the dynamical ages of the 21\,$\mu$m sources to be in the range of 100--1000\,yr.

\cite{McElroy13} theoretically computed the abundance of circumstellar CS based on the UMIST Database for Astrochemistry (UDfA RATE12) network, which incorporates 6173 gas-phase reactions involving 467 atomic and molecular species, as indicated by the solid curve (Model~1) in Fig.~\ref{Figure2}. 
It seems that the observed CS abundance is higher than the model prediction by about one order of magnitude. 
To account for the observations of CS in AGB envelopes, we then adjust the input parameters, including the mass-loss rate, expansion velocity, and initial CS abundance, to 1.4 $\times$ 10$^{-4}$\,$M_{\odot}{\rm yr}^{-1}$, 10\,km\,s$^{-1}$, and 1.0 $\times$ 10$^{-5}$, respectively. 
The updated modeling result is shown by the dashed curve (Model~2) in Fig.~\ref{Figure2}. 
The primary input physical parameters of the CSE models are listed in Table~\ref{Table3}.

According to the chemical model, the abundance of CS would rapidly decrease with the expansion of CSEs. 
This trend is consistent with the observations of CS in AGB envelopes. 
There are 17 reactions (belonging to five kinds of reaction types) related to the consumption of CS in the UDfA RATE12 network \citep{McElroy13}. 
The dominant reactions are photoreactions: 
\begin{align} 
{\rm CS} + h\nu &\longrightarrow {\rm CS}^{+} + {\rm e}^{-}, \\
{\rm CS} + h\nu &\longrightarrow {\rm S} + {\rm C},
\label{Equation4}
\end{align} 
for which the rate coefficients are given by the expression of 
\begin{equation} 
k\,=\,\alpha\,{\rm exp}(-\gamma A_{V})\,{\rm s}^{-1}. 
\label{Equation5}
\end{equation} 
We found that the rate coefficients of reactions~(4) and (5) are 2.05 $\times$ 10$^{-19}$ and 6.30 $\times$ 10$^{-17}$\,s$^{-1}$ at the inner radius of 2.0 $\times$ 10$^{15}$\,cm with a radial extinction $A_{V}$ of 6.9\,mag, while they increase to 1.89 $\times$ 10$^{-10}$ and 9.34 $\times$ 10$^{-10}$ at the outer radius of 7.0 $\times$ 10$^{17}$\,cm with a $A_{V}$ of 0.02\,mag. 
The rate coefficients of other CS destruction reactions are quite similar in the inner and outer envelope regions. 
Therefore, as the CSEs expand, CS can be destroyed by the interstellar UV radiation field with increasing efficiency.

Among the three AGB objects, CRL\,3068 is an extreme carbon star and has the largest $A_{V}$ value. 
\cite{Ramstedt08} showed that the mass-loss rates of CIT\,6, IRC$+$10216, and CRL\,3068 are $0.5$--$1.7 \times 10^{-5}$, $2.0$--$2.1 \times 10^{-5}$, and $2.0$--$6.5 \times 10^{-5}\,M_{\odot}\,{\rm yr}^{-1}$ respectively, with expansion velocities of 17.5, 14.5, and 14.0\,km\,s$^{-1}$. 
We thus obtained the densities of the CSEs, from highest to lowest, in the order of CRL\,3068, IRC$+$10216, and CIT\,6, which are anti-correlated to their CS abundances as shown in Fig.~\ref{Figure2}. 
Therefore, the decrease in the abundance of gas-phase CS in AGB CSEs is likely to be caused by chemical reactions or adsorption onto pre-existing dust grain mantles rather than destruction by the interstellar UV radiation field as the star evolves to the tip of the AGB.

The modeling results are not compatible with the extensive detections of CS in PPNe and PNe. 
The CS abundance of IRAS\,Z02229$+$6208 is similar to that of PNe and significantly lower than that of AGB envelopes, while the gas-phase CSE model predicts that CS could be substantially destroyed during the post-AGB phase. 
During the PPN--PN evolution, CS and other molecules tend to be gradually destroyed in the hardening radiation field from the central star. 
On the other hand, certain molecules could be able to endure the evolutionary phases from PPN and PN to ISM if the CSEs are clumpy. 
\cite{Redman03} simulated the chemical evolution if clumps in circumstellar environments and found that dense clumps can in fact act as a shield to protect some molecules against photodissociation. 
However, the abundance of some short-chain molecules at the late PN stage is still significantly underestimated in the clumpy model. 
The observations of 17 PNe with different ages show that the abundances of HCN and HCO$^{+}$ do not vary as much as those predicted by the clumpy model \citep{Schmidt16}. 
This is also the case for CS \citep{Gold24}. 
According to the model of \cite{Redman03} (see Fig.~\ref{Figure2}), the CS abundance of PNe may drop by three orders of magnitude within $10^4$\,yr, which is not in agreement with the observations. 
Based on these observations, we hypothesize that in certain objects, a protective mechanism differing from the predictions of the clumpy model may operate during the PPN to PN phase, shielding molecules from UV photodestruction.

An inspection of Fig.~\ref{Figure2} shows that the abundances of CS (or their upper limits) in 21\,$\mu$m sources are lower than those in CRL\,2688 and CRL\,618. 
These two objects are PPNe with exceptionally high mass-loss rates, and the resulting high optical depths can effectively shield CS from destruction by UV radiation. 
\cite{Tan24} presented that the sulfur deficit in carbon-rich PNe is likely to be caused by dust chemistry. 
If this were the case, apart from destruction by UV radiation, the variation of CS abundance might be associated with dust chemistry that is more active in the pre-PN phase. 
The envelopes of CRL\,2688 and CRL\,618 have higher density, and presumably, sulfur-bearing molecules can be more effectively adsorbed onto dust grains. 
This contradicts the observations. 
However, shocks may facilitate the release of sulfur-bearing molecules from dust grains back into the gas phase, particularly in sources with energetic outflows. 
High-resolution imaging observations reveal that collimated high-velocity jets characterize the circumstellar environments of CRL\,2688 and CRL\,618 \citep{Cox00, Balick13}, whereas 21\,$\mu$m emission sources are deficient in such jet structures \citep[e.g.,][]{Sun25}. 
This dichotomy offers a plausible alternative explanation for the elevated CS abundance in CRL\,2688 and CRL\,618: 
the shocks driven by these jets can sputter sulfur from dust grain mantles via shock compression and grain-grain collisions, replenishing the gas-phase CS reservoir.

The 30\,$\mu$m emission feature, attributed to MgS dust grains, offers a critical diagnostic for tracing sulfur sequestration in circumstellar environments, potentially illuminating the observed variations in CS abundance. 
Notably, no 30\,$\mu$m emission has been detected in CRL\,2688 or CRL\,618, whereas all 21\,$\mu$m sources exhibit this feature. 
This correlation probably suggests that sulfur is largely depleted into dust grains in 21\,$\mu$m sources, as MgS formation requires efficient condensation of gas-phase sulfur onto refractory carriers. 
The timing of this process is constrained by the fact that 30\,$\mu$m emission is predominantly observed in carbon stars with high dust optical depths, a characteristic of the late AGB phase rather than the PPN stage. 
In contrast, the rapid expansion of PPN envelopes suppresses dust condensation, explaining the absence of 30\,$\mu$m emission in CRL\,2688 and CRL\,618.
However, exceptions exist. 
IRAS\,07134$+$1005 is a 21\,$\mu$m source showing weak 30\,$\mu$m emission yet undetectable CS, challenging the simple depletion scenario. 
This discrepancy may arise from sulfur partitioning into multiple dust phases: 
(1) MgS traced by the 30\,$\mu$m feature; and 
(2) organic sulfur compounds embedded in carbonaceous grains, which do not produce distinct mid-IR features. 
Such complexity underscores the need for multiwavelength observations to fully constrain sulfur chemistry in the CSEs.

\section{Summary}

We present a systematic CS line survey toward a sample of 21\,$\mu$m PPNe. 
While CO emission is detected in all sources, CS is exclusively detected in IRAS\,Z02229+6208, marking the first identification of CS in this object to date. 
Column densities and fractional abundances of CS are derived for all targets. 
Using the UDfA RATE12 chemical reaction network, we model the time evolution of CS fractional abundances, revealing that gas-phase CS is rapidly destroyed within $\sim$10$^3$\,yr after the termination of the AGB phase. 
Contrary to these model predictions, observations indicate that CS can remain abundant in specific PPNe and persist into the PN stage. 
Notably, the CS abundances of 21\,$\mu$m PPNe are significantly lower than those of jet-driven PPNe (e.g., CRL\,2688 and CRL\,618). 
We conclude that three processes likely act synergistically to regulate CS abundance variations across evolutionary stages: 
(1) UV photodissociation, 
(2) shock-driven sulfur release from dust grains in jet-hosting sources, and 
(3) pre-PPN sulfur depletion onto dust grains during the high-mass-loss AGB phase.

This work highlights the chemical diversity of PPNe and underscores the critical role of evolutionary context in shaping circumstellar sulfur chemistry. Future high-sensitivity CS observations with expanded samples, particularly targeting transition objects between the 21\,$\mu$m phase and young PNe, will be essential for constraining the timescales of sulfur reprocessing and validating the multi-process regulatory framework.

\begin{acknowledgments}
The authors would like to express their sincere gratitude to the anonymous reviewer for his/her thorough and constructive comments, which have significantly aided in refining this work.
We wish to express our gratitude to the staff at the SMT 10\,m and PMO 13.7\,m telescopes for their kind help and support during our observations.
The financial supports of this work are from the National Natural Science Foundation of China (NSFC, No. 12463006, 12473027, 12303027, 12333005, and 11973099), the Guangdong Basic and Applied Basic Research Funding (No.\,2024A1515010798), and the science research grants from the China Manned Space Project (NO. CMS-CSST-2021-A09, CMS-CSST-2021-A10, etc). 
Y.Z. thanks the Xinjiang Tianchi Talent Program (2023).
X.D.T. acknowledges the support of the National Key R\&D Program of China under grant No. 2023YFA1608002, the Tianshan Talent Training Program of Xinjiang Uygur Autonomous Region under grant No. 2022TSYCLJ0005, the Chinese Academy of Sciences (CAS) “Light of West China” Program under grant No. xbzg-zdsys-202212, and the Natural Science Foundation of Xinjiang Uygur Autonomous Region under grant No. 2022D01E06.
\end{acknowledgments}

%\vspace{5mm}
\facilities{SMT:10m, PMO:13.7m}
\software{CLASS}

%\bibliography{sample631}{}
%\bibliographystyle{aasjournal}

\appendix
\section{dynamical age}
\label{AppendixA}

For the AGB star CIT\,6, nebular mapping of the CO $J=1 \to 0$ and $2 \to 1$ transitions revealed a radial extent of 25$\arcsec$ \citep{Neri98}. 
The dynamical age of the CSE is determined using the formula $t$ = $D \theta_{\rm env} / V_{\rm exp}$，
where $\theta_{\rm env}$ is the radial angular extent of the envelope, and $V_{\rm exp}$ is its expansion velocity \citep[$16.3$\,km\,s$^{-1}$,][]{Loup93}. 
Using corrected \textit{Gaia} DR3 parallaxes and a Bayesian statistical approach, \cite{Andriantsaralaza22} calculated the distances for approximately 200 AGB stars in the DEATHSTAR\footnote{\url{https://www.astro.uu.se/deathstar/}} project, yielding a distance of 319\,$^{+27}_{-22}$\,pc for CIT\,6. 
Adopting this distance, we derive a dynamical age of 2\,300 $\pm$ 200\,yr.

For the bright AGB star IRC$+$10216, optical observations by \cite{Mauron03} demonstrated that scattered light from its nebula extends up to 200$\arcsec$ from the central star in the \textit{UBV} bands. 
This is consistent with the radio mapping results of the CO $J=1 \to 0$ and $2 \to 1$ emission lines \citep{Huggins88}. 
The DEATHSTAR team derived a distance of $190 \pm 20$\,pc for IRC$+$10216. 
However, they emphasized that the distance obtained via the phase-lag method by \cite{Groenewegen12} is more reliable (i.e., $D=123 \pm 14$\,pc). 
Adopting this distance and the envelope expansion velocity of 14.7\,km\,s$^{-1}$ \citep{Loup93}, we calculated a dynamical age of $8\,500 \pm 970$\,yr.
This value exceeds the dynamical age of CIT\,6 by more than a factor of three. 
We cannot rule out the possibility of bias: IRC$+$10216 is significantly closer to Earth, allowing its more extended outer boundary to be detected.

For the extreme carbon star CRL\,3068, the $V$ band image suggested a radial angular extent of 42$\arcsec$ \citep{Mauron06}. 
The distance of CRL\,3068 given by the DEATHSTAR team is 1\,220\,$^{+70}_{-60}$\,pc. 
Adopting this distance and the envelope expansion velocity of 14.1\,km\,s$^{-1}$ \citep{Loup93}, we estimate the dynamical age to be $17\,000 \pm 1\,000$\,yr. 
It is important to note that the distance to this source is inferred via a Mira-type period-luminosity (P-L) relation. 
This approach introduces two critical uncertainties. 
Firstly, the extinction correction for this object remains poorly constrained. 
More critically, the P-L relation employed was calibrated for pulsation periods spanning 276--514 days, yet CRL\,3068 exhibits a period well outside this empirical range. 
Extrapolating the P-L relation beyond its calibration window can introduce systematic biases. 
Thus, distance determinations for stars with pulsation periods outside the range over which the P-L relation was derived should be treated with caution. 
This caveat applies equally to IRC$+$10216.

For the PPN CRL\,2688, based on an outflow model and the assumption of $D=1$\,kpc, \cite{Jura90} estimated the inner boundary of the dust shell has a radius of $1.4\times10^{16}$\,cm. 
Because the outflow speed along the equatorial direction is 22\,km\,s$^{-1}$, they infer that the intense mass-loss phase ceased about 200 years ago. 
Based on \textit{Hubble Space Telescope (HST)} observations at 2\,$\mu$m band in 1997 and 2002 and 0.6\,$\mu$m band in 2002 and 2009, and assuming an overall self-similar flow for the nebular expansion, \cite{Ueta06} and \cite{Balick12} derived a dynamical age of 350 and 250\,yr, respectively. 
Therefore, we adopt a dynamical age of 200--350\,yr.

For the more evolved PPN CRL\,618, its two visible lobes, which generated by the outflows at the termination of the AGB stage, are separated by about 7$\rlap{.}\arcsec$2 in the [\ion{N}{2}]\,$\lambda6548$ image and have a velocity of 112\,km\,s$^{-1}$ \citep{Carsenty82}. 
Assuming a bolometric luminosity of the central star of 10$^{4}$\,$L_{\odot}$, \cite{Knapp82} and \cite{Knapp85} estimated $D=1.7$ and 1.3\,kpc from IR measurements, respectively. 
It follows that the dynamical age may be 396--518\,yr since this object left the AGB stage.

Optical observations show that the PN K4-47 has two diametrically opposite blobs with a separation of 7$\rlap{.}\arcsec$5 \citep{Corradi00}. 
Adopting an expansion velocity of 150\,km\,s$^{-1}$ \citep{Hartigan87} and a distance ranging from 3--7\,kpc derived through the standard Galactic rotation curve, \cite{Corradi00} obtained a dynamical age of 400--900\,yr. 
The modeling of \cite{Goncalves04} indicated that the separation velocity of the knots is in the range of 250--300\,km\,s$^{-1}$. 
Assuming an expansion velocity of 250\,km\,s$^{-1}$ and a distance of 5.9\,kpc that was derived through a blackbody fitting of the $IRAS$ four-band fluxes \citep{Tajitsu98}, \cite{Edwards14} estimated a dynamical age of about 1\,200\,yr. 
We thus assume a dynamical age of 400--1\,200 yr.

For the PN NGC\,6537, a dust shell formed in a short-lived high mass-loss phase at the very last stage of the AGB star. 
High-resolution images of the \textit{HST} and Very Large Telescope (VLT) suggest that the inner and outer radii of the dust shell are 2$\arcsec$ and $>4\arcsec$, respectively \citep{Matsuura05}. 
The expansion velocity of the interior high-density nebula is about 18\,km\,s$^{-1}$ \citep{Cuesta95, Weinberger89}. 
Through a blackbody fitting to the spectral energy distribution, \cite{Matsuura05} obtained a distance of 0.9--2.4\,kpc. 
Adopting this distance, we estimate a dynamical age of 950--2\,530\,yr.

The [\ion{N}{2}] image of the PN M2-48 shows a pair of collimated lobes with an angular distance of 42$\rlap{.}\arcsec$6 \citep{Manchado96}. 
\cite{Dobrincic08} estimated that the two lobes separated at a velocity of about 175\,km\,s$^{-1}$. 
The average value of the distances given in the literature is 4.2\,kpc, resulting in a dynamical age of 4\,872\,yr \citep{Dobrincic08}. 
Assuming a distance of 1.5--7.7\,kpc, which was derived from the galactic rotation curve, \cite{Edwards14} determined the dynamical age in the range of 1\,800--8\,800\,yr; the value is adopted in this work.

\cite{Harris97} found that the parallax of the PN NGC\,6720 is 1.42 $\pm$ 0.55\,mas. 
Through a three-dimensional model, \cite{ODell07} derived a nebular age of 7\,000\,yr. 
The \textit{Gaia} DR3 indicates a parallax of 1.27 $\pm$ 0.04\,mas and a distance of 787 $\pm$ 25\,pc, resulting in a dynamical age of 7\,800 $\pm$ 300\,yr.

The PN NGC\,6853 (M\,27, the Dumbbell Nebula) exhibits a thick dumbbell-shaped nebula with a half width of 168$\arcsec$ \citep{ODell02} and an expansion velocity of 31\,km\,s$^{-1}$ \citep{Bohuski70}. 
Based on the \textit{Gaia} DR3 parallax of 2.57 $\pm$ 0.04\,mas, we get a trigonometric distance of 389 $\pm$ 6\,pc and a dynamical age of 10\,000 $\pm$ 200\,yr.

For the 21\,$\mu$m source IRAS\,Z02229$+$6208, the spectral energy distribution peaks at about 25\,$\mu$m, which clearly indicates that the maximum dust temperature is significantly lower than that typical for AGB stars \citep{Mishra15}, suggesting the presence of a detached shell. 
Based on the dust inner radius reported by \cite{Mishra16} (corrected using the \textit{Gaia} DR3 distance) and the expansion velocity derived for the two components from our observations (listed in Table~1), we determined the dynamical age of this source during the PPN phase to be 587--954\,yr. 
Based on the Near-IR polarized images of IRAS\,06530$-$0213 and IRAS\,07134$+$1005, \cite{Ueta05} derived the sizes of their inner cavities that were generated by the cessation of mass loss at the end of AGB phase, which are 0$\rlap{.}\arcsec$35--0$\rlap{.}\arcsec$9 and 0$\rlap{.}\arcsec$9--1$\rlap{.}\arcsec$3, respectively. 
The mid-IR images of IRAS\,07134$+$1005 suggested an inner radius of 0$\rlap{.}\arcsec$8--1$\rlap{.}\arcsec$2  for the dust shell \citep{Kwok02}. 
Adopting the \textit{Gaia} DR3 distances of $4.2 \pm 1.2$ and 
$2.2 \pm 0.1$\,kpc,  we get dynamical ages of 440--1\,300\,yr and 900--1\,300\,yr for IRAS\,06530$-$0213 and IRAS\,07134$+$1005 respectively. 
For IRAS\,22272$+$5435, based on mid-infrared images, \cite{Ueta01} suggested that its central star departed the AGB phase approximately 380\,yr ago, following the termination of the superwind, with a distance of 1.6\,kpc. Using the \textit{Gaia} DR3 distance of 1.46 $\pm$ 0.06\,kpc, we derived a dynamical age of 347 $\pm$ 17\,yr. 
The CO interferometric observations of IRAS\,23304$+$6147 showed that the nebula has an equatorial-density-enhancement structure, which is generated during the superwind phase, with an inner diameter of 0$\rlap{.}\arcsec$78 \citep{Sun25}. 
The \textit{Gaia} DR3 parallax is $0.24\pm0.03$\,mas, resulting in a distance of $4.23 \pm 0.06$\,kpc and a dynamical age of $333\pm 5$\,yr. 
Based on the same method used to calculate the dynamical age of IRAS\,Z02229$+$6208 but using the expansion velocities 
reported by \cite{Mishra16}, we estimate that the dynamical ages of IRAS\,05341$+$0852 and IRAS\,22223$+$4327 are 335--590\,yr and 1508--1535\,yr, respectively, since the end of the AGB stage.
It is established that the \textit{Gaia} distances of extended sources may suffer from significant errors, typically overestimating their true distances. 
Therefore, the estimated dynamical ages may represent their upper limits.

\clearpage

\clearpage

\begin{deluxetable*}{ccccccccc}[ht]
\tablecaption{Measurements of the molecular lines, column densities, and fractional abundances.
\label{Table1}}
%\tabletypesize{\scriptsize}
\tablewidth{0pt}
\tablehead{
\colhead{Source}     & \colhead{$V_{\rm exp}$}    & \colhead{$H$}    & \colhead{$T_{\rm p}$}    & \colhead{$\int T_{\rm mb}$d$V$}    & \colhead{RMS}     & \colhead{$V_{\rm LSR}$}     & \colhead{$N_{\rm t}$}     & \colhead{$f$}           \\
\colhead{}                 & \colhead{(km\,s$^{-1}$)}   &                          & \colhead{(mK)}            & \colhead{(mK\,km\,s$^{-1}$)}       & \colhead{(mK)}    & \colhead{(km\,s$^{-1}$)}    & \colhead{(cm$^{-2}$)}  & \colhead{}        
}
\startdata
\hline
\multicolumn{9}{c}{CS $J = 5 \to 4$ Transition} \\
\hline
IRASZ02229$^{a}$        & 10.4 $\pm$ 0.1               & -0.75      & 20.8     & 325.7 $\pm$ 18.7          & 3.9   & 24.6       & (1.6 $\pm$ 0.1) $\times$ 10$^{13}$          & (7.5 $\pm$ 3.8) $\times$ 10$^{-9}$    \\ 
                        & 16.9 $\pm$ 0.1               & -0.66      & 18.8     & 496.2 $\pm$ 24.1          & 3.9   & 21.6       & (2.4 $\pm$ 0.1) $\times$ 10$^{13}$          & (1.1 $\pm$ 0.6) $\times$ 10$^{-8}$    \\ 
IRAS05341               & ...                          & ...        & $<$10    & $<$200                    & 5.3   & ...        & $<$5.3 $\times$ 10$^{13}$                   & $<$8.1 $\times$ 10$^{-7}$             \\ 
IRAS06530               & ...                          & ...        & $<$10    & $<$200                    & 4.7   & ...        & $<$1.6 $\times$ 10$^{13}$                   & $<$2.8 $\times$ 10$^{-8}$              \\ 
IRAS07134               & ...                          & ...        & $<$10    & $<$200                    & 7.0   & ...        & $<$2.8 $\times$ 10$^{13}$                   & $<$1.0 $\times$ 10$^{-8}$              \\ 
IRAS22223               & ...                          & ...        & 15:      & $<$200                    & 6.8   & ...        & $<$5.3 $\times$ 10$^{13}$                   & $<$8.4 $\times$ 10$^{-9}$              \\ 
IRAS22272               & ...                          & ...        & $<$10    & $<$200                    & 5.0   & ...        & $<$5.2 $\times$ 10$^{13}$                   & $<$3.6 $\times$ 10$^{-9}$              \\ 
IRAS23304               & ...                          & ...        & $<$10    & $<$200                    & 2.8   & ...        & $<$9.1 $\times$ 10$^{13}$                   & $<$6.6 $\times$ 10$^{-9}$              \\ 
\hline
\multicolumn{9}{c}{CO $J = 1 \to 0$ Transition} \\
\hline
IRAS05341               & 7.3 $\pm$ 1.4                & -0.9       & 28.7     & 290.6 $\pm$ 34.3          & 7.8   & 6.4         & (1.7 $\pm$ 0.2) $\times$ 10$^{17}$         & (1.1 $\pm$ 0.5) $\times$ 10$^{-3}$      \\ 
IRAS22223               & 10.6 $\pm$ 0.1               & -0.3       & 70.0     & 1314.8 $\pm$ 45.6         & 11.3  & -31.3       & (6.7 $\pm$ 0.3) $\times$ 10$^{16}$         & (4.8 $\pm$ 2.4) $\times$ 10$^{-5}$      \\ 
\hline
\hline
\enddata
\tablecomments{
\tablenotetext{a}{Derived from the broad and narrow components of the CS $J = 5 \to 4$ transition, respectively.}
}
\end{deluxetable*}

\begin{deluxetable*}{lccccc}
\tablecaption{Fractional abundances of CS and dynamical ages of other CSEs.
\label{Table2}}
\tabletypesize{\normalsize}
\tablewidth{0pt}
\tablehead{
\colhead{Objects}       & \colhead{$f$(CS)}     & \colhead{Reference}   & \colhead{Age (yr)$^{a}$}     & \colhead{Type}     
}
\startdata
CIT\,6             & 1.98 $\times$ 10$^{-6}$                          & 1    & 2\,300 $\pm$ 200        & AGB         \\
                   & 2.0 $\times$ 10$^{-6}$                           & 2    &                         &             \\
                   & 2.5 $\times$ 10$^{-6}$                           & 3    &                         &             \\
                   & 7.0 $\times$ 10$^{-6}$                           & 4    &                         &             \\
IRC$+$10216        & 9.9 $\times$ 10$^{-7}$                           & 3    & 12\,000 $\pm$ 1\,000    & AGB         \\
                   & 4.6 $\times$ 10$^{-7}$                           & 3    &                         &             \\
CRL\,3068          & 5.8 $\times$ 10$^{-8}$                           & 5    & 17\,000 $\pm$ 1\,000    & AGB         \\
                   & 3.7 $\times$ 10$^{-7}$                           & 3    &                         &             \\
CRL\,2688          & 1.64 $\times$ 10$^{-7}$                          & 6    & 275 $\pm$ 75            & PPN         \\
                   & 2.0 $\times$ 10$^{-7}$                           & 7    &                         &             \\
                   & 2.1 $\times$ 10$^{-6}$                           & 8    &                         &             \\
CRL\,618           & 6.0 $\times$ 10$^{-8}$                           & 7    & 457 $\pm$ 61            & PPN         \\
                   & 4.1 $\times$ 10$^{-7}$                           & 8    &                         &             \\
K4-47              & (1.9 $\pm$ 1.3) $\times$ 10$^{-8}$               & 9    & 1\,475 $\pm$ 495        & PN          \\
NGC\,6537            & (4.3 $\pm$ 0.5 $\times$ 10$^{-8}$                & 9    & 1\,740 $\pm$ 790        & PN          \\
M2-48              & (2.7 $\pm$ 1.5) $\times$ 10$^{-8}$               & 9    & 5\,300 $\pm$ 3\,500     & PN          \\
NGC\,6720          & (1.89 $\pm$ 0.97) $\times$ 10$^{-8}$             & 9    & 7\,800 $\pm$ 300        & PN          \\
NGC\,6853          & (1.24 $\pm$ 0.52) $\times$ 10$^{-8}$             & 9    & 10\,000 $\pm$ 200       & PN          \\
\hline
\hline
\enddata
\tablecomments{
\tablenotetext{a}{Dynamical ages of the PPNe and PNe are counted starting at
 the termination of the AGB stage.}
}
{\bf Reference.} {(1) \cite{Yang23}; (2) \cite{Zhang09a}; (3) \cite{Woods03}; (4) \cite{Chau12}; (5) \cite{Zhang09b}; (6) \cite{Qiu22}; (7) \cite{Bujarrabal88}; (8) \cite{Bujarrabal94}; (9) \cite{Edwards14}.}
\end{deluxetable*}

\begin{deluxetable*}{lccccc}
\tablecaption{Primary input physical parameters of the CSE models.
\label{Table3}}
\tabletypesize{\normalsize}
\tablewidth{0pt}
\tablehead{
\colhead{Parameters}       & \colhead{Model~1}       & \colhead{Model~2}         & \colhead{Unit}  
}
\startdata
Initial CS abundance ($f_{\rm CS}$)         & 7.0 $\times$ 10$^{-7}$        & 1.0 $\times$ 10$^{-5}$   & \nodata                      \\
Mass loss rate ($\dot{M}_{\rm H_{2}}$)      & 1.5 $\times$ 10$^{-5}$        & 1.4 $\times$ 10$^{-4}$   & $M_{\odot}\,{\rm yr}^{-1}$   \\
Expansion velocity ($V_{\rm exp}$)          & 14.5                          & 10.0                     & km\,s$^{-1}$                 \\
Initial radius ($r_{i}$)                    & 2.0 $\times$ 10$^{15}$        & 2.0 $\times$ 10$^{15}$   & cm                           \\
Initial kinetic temperature ($T$)           & 221                           & 221                      & K                            \\
Initial H$_{2}$ density ($n_{\rm H{_2}}$)   & 3.2 $\times$ 10$^{-6}$        & 3.2 $\times$ 10$^{-6}$   & cm$^{-3}$                    \\
Initial dust extinction ($A_{V}$)           & 6.9                           & 6.9                      & mag                          \\
\hline
\hline
\enddata
\tablecomments{
The input physical parameters of Model~1 are the same as the CSE model of \cite{McElroy13}. 
}
\end{deluxetable*}

\end{CJK*}
\end{document}